\documentclass[11pt]{article}

\usepackage[T1]{fontenc}
\usepackage[utf8]{inputenc}
\usepackage{lmodern}
\usepackage[a4paper,margin=1in]{geometry}

\usepackage{graphicx}
\usepackage{xspace}
\usepackage{microtype}
\usepackage{booktabs}
\usepackage{longtable}
\usepackage{tabularx}
\usepackage{array}
\usepackage{listings}
\usepackage{xcolor}
\usepackage{url}
\usepackage{tikz}
\usepackage{pifont}
\usepackage{hyperref}

\usetikzlibrary{arrows.meta,positioning,fit,backgrounds,calc}

\newcommand{\prodock}{\texttt{ProDock}\xspace}
\newcommand{\pkg}[1]{\texttt{#1}\xspace}
\newcommand{\class}[1]{\texttt{#1}\xspace}
\newcommand{\code}[1]{\texttt{#1}\xspace}

\lstdefinelanguage{json}{
    basicstyle=\ttfamily\small,
    showstringspaces=false,
    breaklines=true,
    breakatwhitespace=true,
    morestring=[b]",
    stringstyle=\color{black},
    literate=
     *{0}{{{\color{black}0}}}{1}
      {1}{{{\color{black}1}}}{1}
      {2}{{{\color{black}2}}}{1}
      {3}{{{\color{black}3}}}{1}
      {4}{{{\color{black}4}}}{1}
      {5}{{{\color{black}5}}}{1}
      {6}{{{\color{black}6}}}{1}
      {7}{{{\color{black}7}}}{1}
      {8}{{{\color{black}8}}}{1}
      {9}{{{\color{black}9}}}{1}
      {:}{{{\color{black}{:}}}}{1}
      {,}{{{\color{black}{,}}}}{1}
      {\{}{{{\color{black}{\{}}}}{1}
      {\}}{{{\color{black}{\}}}}}{1}
      {[}{{{\color{black}{[}}}}{1}
      {]}{{{\color{black}{]}}}}{1},
}

\definecolor{jsonbg}{RGB}{247,248,250}
\definecolor{jsonline}{RGB}{90,90,90}

\definecolor{pdline}{RGB}{70,70,70}
\definecolor{pdroot}{RGB}{236,239,243}
\definecolor{pdmain}{RGB}{245,246,248}
\definecolor{pdpre}{RGB}{228,236,248}
\definecolor{pdpost}{RGB}{231,241,233}
\definecolor{pdsub}{RGB}{252,252,252}

\hypersetup{
  colorlinks=true,
  linkcolor=blue,
  citecolor=blue,
  filecolor=blue,
  urlcolor=blue,
  anchorcolor=blue,
  pdfauthor={Tieu-Long Phan and coauthors},
  pdftitle={ProDock: From multi-target consensus docking into database-backed storage}
}

\title{\textbf{ProDock: From multi-target consensus docking into database-backed storage}}

\author{
Tieu-Long Phan$^{1,2}$\thanks{Corresponding authors: \href{mailto:tieu@bioinf.uni-leipzig.de}{tieu@bioinf.uni-leipzig.de}; \href{mailto:truongtuyen@ump.edu.vn}{truongtuyen@ump.edu.vn}} \and
Lai Hoang Son Le$^{3}$ \and
Thanh-An Pham$^{1}$ \and
Nhu-Ngoc Nguyen Song$^{4}$ \and
Tuyet-Minh Phan$^{1,5}$ \and
Tuyen Ngoc Truong$^{4}$\footnotemark[1]
}

\date{
\small
$^{1}$ Bioinformatics Group, Department of Computer Science \& Interdisciplinary Center for Bioinformatics \& School for Embedded and Composite Artificial Intelligence (SECAI), Leipzig University, Leipzig, Germany\\
$^{2}$ Department of Mathematics and Computer Science, University of Southern Denmark, Odense M, Denmark\\
$^{3}$ Department of Chemistry, Texas A\&M University, College Station, TX 77843, USA\\
$^{4}$ School of Pharmacy, University of Medicine and Pharmacy at Ho Chi Minh City, Ho Chi Minh City, Vietnam\\
$^{5}$ Department of Theoretical Chemistry, University of Vienna, Vienna, Austria\\[1em]
}

\begin{document}

\maketitle

\begin{abstract}
Protein--ligand docking is widely used in structure-based discovery, but routine studies often fail at the workflow level rather than at the scoring level. Receptor cleaning, ligand preparation, file conversion, box definition, run organization, and downstream parsing are frequently handled by fragmented scripts, which reduces reproducibility, obscures provenance, and complicates comparative analysis across targets, ligands, and docking settings. We present \prodock, an open-source Python toolkit for reproducible protein--ligand docking and postprocessing. \prodock organizes application-oriented docking into four connected layers: receptor and ligand preprocessing, provenance-aware docking execution, postprocessing of poses and interaction fingerprints, and SQLite-backed storage for later querying. The package supports inputs ranging from PDB identifiers and local receptor files to \texttt{SMILES} strings and prepared ligand directories, and integrates receptor preparation, ligand preparation, reference-ligand-based box generation, campaign serialization, batch docking, pose crawling, score extraction, interaction profiling, and database insertion within a consistent project-local workflow. By representing studies as explicit many-to-many campaigns linking multiple receptors, ligands, and docking backends, \prodock converts fragmented engine-specific outputs into structured analytical results that are easier to compare, reuse, and audit. \prodock is implemented in Python and released under an open-source license at \url{https://github.com/Medicine-Artificial-Intelligence/ProDock}. Documentation is available at \url{https://prodock.readthedocs.io/en/latest}.
\end{abstract}


\section{Introduction}

Structure-based docking remains a widely used technique in computer-aided drug discovery because it offers fast and testable hypotheses for ligand binding modes and compound prioritization~\cite{bender2021large}. In practice, however, the scientific value of docking depends not only on the search algorithm or scoring function but also on the surrounding workflow used to prepare receptors, generate ligand structures, define binding regions, execute calculations, and organize outputs for later inspection. A routine docking study often requires retrieval or cleaning of receptor structures, extraction of reference ligands or cofactors, generation of three-dimensional ligand coordinates from \texttt{SMILES} strings, handling of protonation and hydrogens, conversion among file types such as \texttt{SDF}, \texttt{PDB}, and \texttt{PDBQT}, selection of search boxes, execution of repeated runs, and transformation of raw outputs into forms suitable for ranking and structural interpretation.

These steps are essential, yet they are still frequently handled through small project-specific scripts that differ across laboratories, datasets, and computing environments. As a result, minor inconsistencies in receptor or ligand preparation can propagate into failed runs or hard-to-reproduce results. Intermediate files are often generated without clear records of the parameters that created them, and postprocessing is commonly separated from docking itself, forcing users to recover scores, poses, and interaction summaries with one-off parsers. Established packages such as \texttt{RDKit}~\cite{rdkit} and \texttt{Open Babel}~\cite{oboyle2011openbabel} provide robust functionality for cheminformatics and structure conversion, while members of the \texttt{AutoDock Vina} family, including \texttt{Smina}~\cite{koes2013smina}, \texttt{QuickVina~2}~\cite{Alhossary2015QVina}, and \texttt{GNINA}~\cite{mcnutt2021gnina}, provide the search and scoring machinery used in practical screening. Workflow-oriented frameworks such as \texttt{DockStream}~\cite{Guo2021DockStream}, \texttt{EasyDock}~\cite{Minibaeva2023EasyDock}, and \texttt{pyscreener}~\cite{Graff2022pyscreener} have improved reproducibility, but many application-focused studies still depend on fragmented preparation, execution, and analysis steps that are difficult to audit or reuse.

\prodock was developed to address this application-level gap. Rather than proposing a new scoring model, the package provides a lightweight Python framework that integrates four connected stages of a docking study: preprocessing of receptors and ligands, provenance-aware execution of docking campaigns, postprocessing of poses and interaction fingerprints, and persistent storage of structured results in a project-local SQLite database. This design targets application-oriented studies in which reproducibility, traceability, and downstream analysis are as important as the docking calculation itself.

A central design principle of \prodock is that each stage should remain explicit, reusable, and easy to inspect. Prepared inputs are retained rather than hidden inside transient scripts, campaign parameters are serialized in machine-readable form, and postprocessing yields table-oriented outputs that can be queried directly or inserted into a local database. By enforcing a stable project layout and preserving validated intermediate artifacts, \prodock reduces the practical overhead that often limits replication of docking workflows across datasets and collaborators. The software is distributed through \texttt{PyPI} and \texttt{conda}, with installation and dependency details described in Supporting Section~\ref{sec:si-software}.

\section{Implementation}

\begin{figure}[htbp]
\centering
\IfFileExists{Figure/prodock_flow.pdf}{%
    \includegraphics[width=\linewidth]{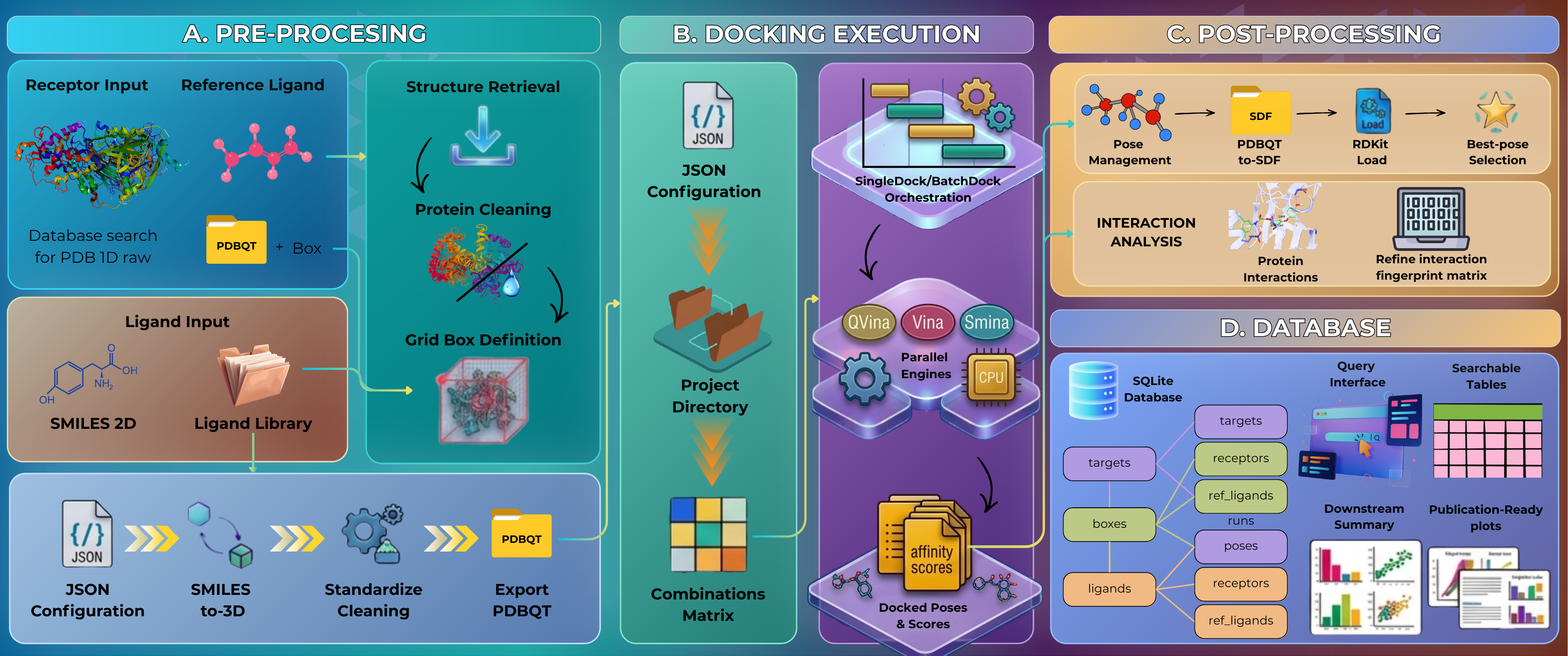}%
}{%
    \fbox{\parbox[c][6.2cm][c]{0.95\linewidth}{\centering
    Placeholder for Figure 1. Recommended structure: a four-stage workflow diagram showing preprocessing of receptor and ligand inputs, provenance-aware docking execution, postprocessing of poses and interaction fingerprints, and SQLite-backed storage for downstream querying and comparison.}}%
}
\caption{Overview of the \prodock workflow. The package organizes docking into four connected stages: preprocessing of receptor and ligand inputs, docking execution, postprocessing of poses and interaction fingerprints, and SQLite result repository for downstream querying and comparison.}
\label{fig:prodock-overview}
\end{figure}

\prodock is organized into a four-stage workflow, including \emph{preprocessing}, \emph{docking}, \emph{postprocessing}, and \emph{database storage} (see Figure~\ref{fig:prodock-overview}), and its architecture is divided into five subpackages, detailed in Supporting Section~\ref{sec:si-software}.

\subsection{Preprocessing of receptors and ligands}

The workflow begins with preprocessing (see Figure~\ref{fig:prodock-overview}A), where raw receptor and ligand inputs are converted into consistent docking-ready representations. Receptors may originate from a PDB identifier or from local structure files. When a target is obtained through \texttt{PDBQuery}, \prodock can standardize the corresponding local representation by retrieving the structure, filtering chains, retaining selected cofactors, and extracting a reference ligand when available. This step makes receptor definition explicit and reproducible before docking begins.

Receptor preparation is handled by \texttt{ReceptorPrep}, 
which cleans and refines the structural model before docking. Depending on the input, this stage can repair common structural issues, remove undesired heterogens, add missing atoms and hydrogens, and optionally minimize coordinates. These tasks rely on \texttt{PDBFixer}~\cite{eastman2017openmm7} together with optional minimization through \texttt{OpenMM}~\cite{eastman2023openmm8} or \texttt{Open Babel}~\cite{oboyle2011openbabel}. The prepared receptor is then converted into docking-compatible \texttt{PDBQT} using \texttt{Open Babel} or \texttt{Meeko}~\cite{martins2025meeko}. Because the preparation stage is explicit rather than hidden within engine-specific wrappers, the resulting receptor files remain reusable across repeated campaigns and straightforward to inspect when troubleshooting is needed.

Ligand preparation is handled separately by \texttt{LigandPrep}, which accepts \texttt{SMILES} strings, tabular records, mapping-style inputs, or previously prepared ligand directories. Using \texttt{RDKit}~\cite{rdkit} as the primary cheminformatics backend, \prodock can generate three-dimensional ligand structures, add hydrogens, embed conformers, optimize coordinates, and export ligands into formats required for docking. This separation between ligand generation and docking is important in practice because the same curated library may be screened against multiple receptors, and reused for method comparison or selection. Keeping ligand preparation explicit, therefore, improves both reproducibility and efficiency.

Definition of the search region is integrated into the same preprocessing layer through the dedicated \texttt{GridBox} module. Users may specify box coordinates directly, but \prodock can also infer the docking box from a reference ligand, which reflects common structure-based workflows based on co-crystallized binders. Further details regarding the autobox algorithm are provided in Supporting Section~\ref{sec:si-gridbox}. Integrating receptor preparation, ligand preparation, and box generation within one stage reduces the risk of silent failures caused by malformed structures, missing hydrogens, or poorly defined search regions. In this sense, preprocessing is not a preliminary convenience step; it is a core part of scientific quality control.

\subsection{Docking execution}

Once receptors, ligands, and search regions have been prepared, \prodock organizes docking as a structured campaign rather than as a collection of isolated command-line calls (see Figure~\ref{fig:prodock-overview}B). In the current implementation, individual jobs are managed through \texttt{SingleDock}, whereas larger screening workflows are coordinated through \texttt{BatchDock}. Multiple docking backends are supported through a common execution interface, allowing the surrounding workflow to remain stable even when the selected engine changes.

A key idea in this stage is normalization of backend-specific execution details. Rather than exposing engine-specific syntax at the user level, \prodock translates prepared receptors, ligands, search boxes, and runtime options into standardized task definitions. These tasks can then be dispatched consistently across supported engines while preserving the same logical campaign structure. This abstraction allows users to work at the study level without losing access to engine-specific parameters.

For transparency and reproducibility, each study is recorded as a machine-readable JSON campaign file organized around receptors. A single campaign can include multiple receptors, each linked to its prepared structure, docking box, ligand collections, and engine-specific execution settings. The resulting data model therefore supports many-to-many study designs in which multiple receptors are evaluated against multiple ligands and, when desired, across multiple docking backends within the same campaign. By preserving key runtime parameters such as CPU allocation, random seed, exhaustiveness, and requested pose count, the campaign file functions both as an execution plan and as a portable provenance record.

This execution model is designed for practical screening workflows. Jobs can be run in parallel, while output poses and score logs are written into a structured project hierarchy. As a result, \prodock reduces the need for ad hoc orchestration scripts and makes docking results easier to organize, compare, audit, and reproduce across studies of moderate scale. The execution layer also establishes a stable bridge between preparation and analysis, because the same campaign description can be used to understand how a result was produced after the computation has already finished.

\subsection{Postprocessing of poses, scores, and interaction fingerprints}

The third stage of the workflow is postprocessing (see Figure~\ref{fig:prodock-overview}C), where raw docking outputs are transformed into structured data for downstream analysis. Rather than leaving users with scattered pose files and engine-specific text logs, \prodock organizes this stage around a dedicated processing layer centered on \texttt{PoseCrawler}. This component can recover docked poses from a single output file, a flat directory, or a full project hierarchy, making it suitable for both exploratory calculations and larger screening campaigns. By extracting pose identifiers, ranks, engines, and scores into a common tabular representation, \prodock standardizes heterogeneous outputs and reduces the amount of manual inspection required after docking.

Postprocessing also extends beyond score recovery by incorporating protein--ligand interaction fingerprinting. Because numerical ranking alone is rarely sufficient to interpret docking outcomes, the \texttt{InteractionProfiler} module integrates \texttt{ProLIF}~\cite{bouysset2021prolif} to automate receptor loading, ligand handling, and residue-level interaction analysis. This enables users to examine whether highly ranked poses preserve chemically meaningful contacts, reproduce known recognition patterns, or differ from related ligands in interpretable ways. Integrating interaction analysis directly into the workflow makes structural interpretation more reproducible and easier to compare across campaigns. 

To support different analysis needs, \prodock generates both compact pose-level summaries and expanded interaction tables. Compact summaries are useful for filtering, aggregation, and large-scale comparison, whereas expanded event tables preserve the residue-level detail needed for close structural inspection. In this way, postprocessing becomes an analytical stage rather than a simple reporting step, converting raw docking outputs into data structures that are ready for cheminformatics analysis, biological interpretation, and database storage.

\subsection{Database storage and querying}

The final stage of the workflow is persistent storage, as illustrated in Figure~\ref{fig:prodock-overview}D. After preprocessing, docking, and postprocessing have generated structured pose and interaction data, \prodock can consolidate these results in a dedicated SQLite database through components such as \texttt{PoseDatabase} and related query utilities. This step turns a completed docking run into a reusable project resource rather than a transient set of output files.

Database-backed storage is especially useful for medium-scale and larger studies, where the bottleneck often shifts from generating outputs to retrieving and comparing them efficiently. The schema organizes receptors, ligands, engines, poses, scores, and interaction records in a form that can be queried by receptor, ligand, backend, pose rank, score threshold, or residue-level interaction pattern. This separation of calculation from analysis allows users to execute docking once and then perform ranking, reporting, filtering, and interpretation repeatedly without reparsing raw files.

SQLite is a pragmatic choice because it fits the local and portable design of \prodock. It requires no separate server infrastructure and can be archived together with the campaign JSON, prepared inputs, and derived outputs. This storage model also reinforces the many-to-many organization of a campaign, because receptors, ligands, engines, poses, and interactions can all be linked in a queryable form within one project. As a result, the database is not merely a convenience for archiving; it is a practical analytical layer that supports comparison across targets, ligands, and docking configurations. Supporting Section~\ref{sec:si-db} provides further details on the database architecture and query utilities.

\section{Typical use}

A typical \prodock workflow begins with a receptor provided either as a PDB identifier or as a local structure file, together with a ligand set defined from \texttt{SMILES} strings, tabular records, or prepared ligand files. The package then prepares receptors and ligands, defines the search region from explicit box coordinates or a reference ligand, and writes a serialized campaign file that records the selected docking engines and runtime settings. The resulting campaign can be executed through either the Python API or the command-line interface (see Supporting Section~\ref{sec:si-usage}).

\begin{lstlisting}[language=bash,caption={Illustrative command-line execution of a \prodock campaign.}]
python -m prodock --config config.json --receptor-json receptor.json --ligand-json ligand.json
\end{lstlisting}

In this workflow, \code{config.json} stores execution
settings, whereas \code{receptor.json} and \code{ligand.json} define the molecular inputs. After docking, \prodock recovers ranked poses, extracts scores, optionally computes protein--ligand interaction fingerprints, and stores structured outputs in a local SQLite database for downstream analysis. Because the workflow is modular, users may adopt the full end-to-end pipeline or reuse only specific components such as receptor preparation, pose crawling, or database insertion within existing projects. This flexibility is important for adoption, since many groups already possess partial workflows and need a tool that integrates with them rather than forcing a complete replacement. This architecture also simplifies teaching, benchmarking, maintenance, and collaborative reuse.

To demonstrate the utility of \prodock, we conducted an \textit{in silico} screening campaign against EGFR, targeting five distinct crystal structures (\texttt{2ITY}, \texttt{1M17}, \texttt{4G5J}, \texttt{4I23}, and \texttt{4ZAU}) using a library of established inhibitors and decoy molecules. Provided solely with PDB identifiers and ligand \texttt{SMILES} strings, \prodock autonomously executed the necessary preparatory steps: curating 3D receptor and ligand structures, defining search spaces based on co-crystallized reference ligands, and orchestrating the docking simulations across four distinct search engines (\texttt{smina}, \texttt{vina}, \texttt{qvina}, and \texttt{qvina-w}). This highlights the primary advantage of \prodock for computational chemists: consolidating complex, multi-target docking studies from a collection of disparate ad hoc scripts into a streamlined, highly reproducible workflow. For comprehensive computational parameters and workflow configurations, see Supporting Section~\ref{sec:si-case-study}.

\section{Conclusion}

\prodock addresses a practical gap in structure-based docking by connecting receptor and ligand preparation, provenance-aware docking execution, postprocessing, and persistent result storage within a single lightweight framework. Rather than proposing a new docking algorithm, it focuses on the workflow infrastructure that often determines whether docking studies are reproducible, traceable, and straightforward to analyze. This makes the package particularly useful for application-oriented projects in which multiple receptors, ligand libraries, and docking settings must be handled consistently within one study.

A particular strength of \prodock is that it supports docking as a structured many-to-many workflow rather than as a collection of isolated runs. A single campaign can associate multiple receptors with multiple ligands and multiple docking backends, while the accompanying SQLite schema preserves these relationships together with poses, scores, and interaction fingerprints in a queryable form. By combining pose recovery, score extraction, interaction profiling, and database-backed storage, \prodock converts scattered engine-specific outputs into structured analytical results that are easier to compare, filter, and reuse across screening and benchmarking studies.

At the same time, \prodock is intentionally modest in scope. It does not seek to replace established docking engines or to overcome the intrinsic limitations of docking accuracy and scoring. Instead, it provides a practical and extensible software layer that helps these tools be used more systematically and their outputs be analyzed more effectively. By consolidating fragmented preparation, execution, and analysis steps into a coherent workflow, \prodock provides a reproducible foundation for routine docking campaigns and future workflow-oriented development.

\section*{Competing interests}
No competing interest is declared.

\section*{Author contributions}
T.-L.P. conceived the framework, designed the software and drafted the manuscript. N.-N.S.N., T.-M.P. and L.H.S.L. contributed to implementation, testing and application design. T.N.T. supervised the study and contributed to manuscript revision. All authors read and approved the final manuscript.

\section*{Funding}
This work has received support from the Korea International Cooperation Agency (KOICA) under the project entitled ``Education and Research Capacity Building Project at University of Medicine and Pharmacy at Ho Chi Minh City'', conducted from 2025 to 2026 (Project No.~2021-00020-3).

\section*{Data availability}
The source code, documentation and example datasets are available from the \prodock project repository at \url{https://github.com/Medicine-Artificial-Intelligence/ProDock}.

\clearpage

% =========================
% Supporting Information
% =========================

\setcounter{page}{1}
\renewcommand{\thepage}{S\arabic{page}}

\setcounter{section}{0}
\setcounter{subsection}{0}
\setcounter{subsubsection}{0}
\setcounter{figure}{0}
\setcounter{table}{0}
\setcounter{equation}{0}

\renewcommand{\thesection}{S\arabic{section}}
\renewcommand{\thesubsection}{S\arabic{section}.\arabic{subsection}}
\renewcommand{\thesubsubsection}{S\arabic{section}.\arabic{subsection}.\arabic{subsubsection}}
\renewcommand{\thefigure}{S\arabic{figure}}
\renewcommand{\thetable}{S\arabic{table}}
\renewcommand{\theequation}{S\arabic{equation}}

\section*{Supporting Information}
\addcontentsline{toc}{section}{Supporting Information}

\section{Software overview}
\label{sec:si-software}

\subsection{Software architecture}

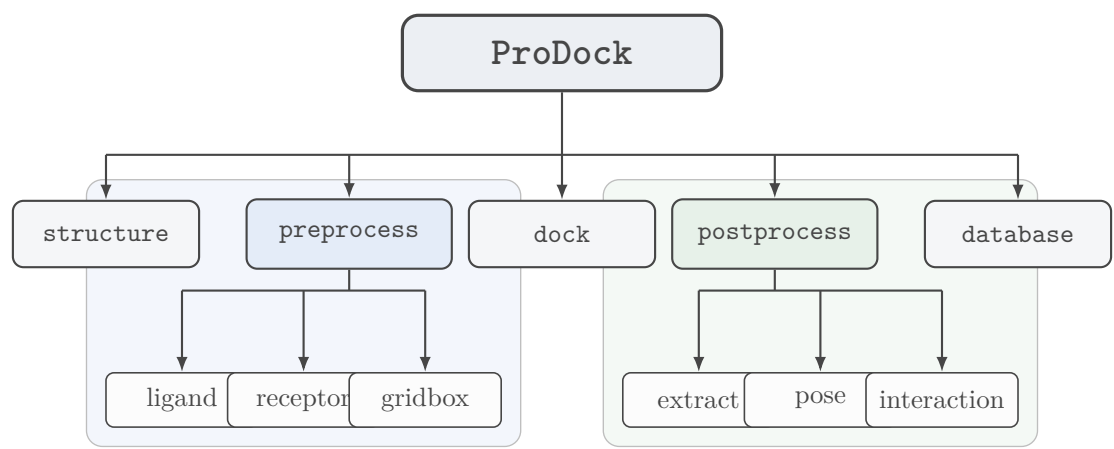
\begin{figure}[htbp]
\centering
\begin{tikzpicture}[
    font=\small,
    >=Latex,
    text=pdline,
    root/.style={
        draw=pdline,
        very thick,
        rounded corners=6pt,
        minimum width=4.2cm,
        minimum height=1.0cm,
        align=center,
        fill=pdroot
    },
    main/.style={
        draw=pdline,
        thick,
        rounded corners=4pt,
        minimum width=2.45cm,
        minimum height=0.88cm,
        align=center,
        fill=pdmain
    },
    prebox/.style={
        draw=pdline,
        thick,
        rounded corners=4pt,
        minimum width=2.7cm,
        minimum height=0.92cm,
        align=center,
        fill=pdpre
    },
    postbox/.style={
        draw=pdline,
        thick,
        rounded corners=4pt,
        minimum width=2.7cm,
        minimum height=0.92cm,
        align=center,
        fill=pdpost
    },
    sub/.style={
        draw=pdline,
        semithick,
        rounded corners=3pt,
        minimum width=2.0cm,
        minimum height=0.72cm,
        align=center,
        fill=pdsub
    },
    conn/.style={draw=pdline, thick},
    arr/.style={draw=pdline, thick, -{Latex[length=2mm]}}
]

\node[root] (prodock) at (0,0) {\Large\textbf{\prodock}};

\node[main]   (structure)   at (-6.0,-2.4) {\pkg{structure}};
\node[prebox] (preprocess)  at (-2.8,-2.4) {\pkg{preprocess}};
\node[main]   (dock)        at ( 0.0,-2.4) {\pkg{dock}};
\node[postbox](postprocess) at ( 2.8,-2.4) {\pkg{postprocess}};
\node[main]   (database)    at ( 6.0,-2.4) {\pkg{database}};

\coordinate (t1) at (0,-0.75);
\coordinate (t2) at (0,-1.35);
\draw[conn] (prodock.south) -- (t1) -- (t2);
\draw[conn] (-6.0,-1.35) -- (6.0,-1.35);

\draw[arr] (-6.0,-1.35) -- (structure.north);
\draw[arr] (-2.8,-1.35) -- (preprocess.north);
\draw[arr] ( 0.0,-1.35) -- (dock.north);
\draw[arr] ( 2.8,-1.35) -- (postprocess.north);
\draw[arr] ( 6.0,-1.35) -- (database.north);

\node[sub] (ligand)   at (-5.0,-4.6) {ligand};
\node[sub] (receptor) at (-3.4,-4.6) {receptor};
\node[sub] (gridbox)  at (-1.8,-4.6) {gridbox};

\coordinate (p1) at (-2.8,-3.15);
\draw[conn] (preprocess.south) -- (p1);
\draw[conn] (-5.0,-3.15) -- (-1.8,-3.15);
\draw[arr] (-5.0,-3.15) -- (ligand.north);
\draw[arr] (-3.4,-3.15) -- (receptor.north);
\draw[arr] (-1.8,-3.15) -- (gridbox.north);

\node[sub] (extract)     at (1.8,-4.6) {extract};
\node[sub] (pose)        at (3.4,-4.6) {pose};
\node[sub] (interaction) at (5.0,-4.6) {interaction};

\coordinate (q1) at (2.8,-3.15);
\draw[conn] (postprocess.south) -- (q1);
\draw[conn] (1.8,-3.15) -- (5.0,-3.15);
\draw[arr] (1.8,-3.15) -- (extract.north);
\draw[arr] (3.4,-3.15) -- (pose.north);
\draw[arr] (5.0,-3.15) -- (interaction.north);

\begin{scope}[on background layer]
\node[
    rounded corners=6pt,
    fill=pdpre!45,
    draw=pdline!35,
    line width=0.5pt,
    inner sep=7pt,
    fit=(preprocess)(ligand)(receptor)(gridbox)
] {};
\node[
    rounded corners=6pt,
    fill=pdpost!45,
    draw=pdline!35,
    line width=0.5pt,
    inner sep=7pt,
    fit=(postprocess)(extract)(pose)(interaction)
] {};
\end{scope}

\end{tikzpicture}
\caption{Hierarchical organization of the \prodock package. The package root expands into the core modules \pkg{structure}, \pkg{preprocess}, \pkg{dock}, \pkg{postprocess}, and \pkg{database}. The \pkg{preprocess} module contains ligand, receptor, and grid-box preparation, whereas the \pkg{postprocess} module contains extraction, pose analysis, and interaction profiling.}
\label{fig:prodock-architecture}
\end{figure}

\prodock follows a modular software design centered on a four-stage workflow comprising preprocessing, docking with provenance tracking, postprocessing, and database-backed analysis. As illustrated in Figure~\ref{fig:prodock-architecture}, this workflow is reflected directly in the package structure, in which distinct modules are responsible for structure handling, input preparation, docking execution, result parsing, and persistent storage. In particular, target and ligand preparation are implemented in \pkg{prodock.preprocess}, docking orchestration is handled by \pkg{prodock.dock}, structured downstream analysis is provided by \pkg{prodock.postprocess}, and long-term result storage together with query utilities is implemented in \pkg{prodock.database}. Supporting modules such as \pkg{prodock.structure} further separate low-level structure retrieval and manipulation from higher-level workflow execution.

At the framework level, \class{ProDock} brings these modules together within a project-local execution model that preserves both intermediate artifacts and final outputs in a consistent directory layout. This design supports reproducible end-to-end docking campaigns while retaining the modularity needed to reuse individual stages independently. The same separation of concerns is reflected in the command-line interface, which accepts either a unified all-in-one JSON specification or split JSON inputs for receptors, ligands, and campaign settings. In practice, this flexibility supports both rapid single-command execution and more iterative workflows in which prepared inputs or campaign definitions are reused across repeated experiments.

\subsection{Installation and availability}
\label{sec:si-install}

\prodock is distributed through both PyPI and Conda, with documentation available at \url{https://prodock.readthedocs.io/en/latest/}. For users who prefer a single-step installation, the Conda package is the most convenient option because it installs \prodock together with required dependencies. A representative installation command is:

\begin{lstlisting}[language=bash,caption={Recommended installation using Conda.}]
conda install -c tieulongphan prodock
\end{lstlisting}

Alternatively, \prodock can be installed from PyPI:

\begin{lstlisting}[language=bash,caption={Installation from PyPI.}]
pip install prodock
\end{lstlisting}

\prodock is built for Python 3.11 to maintain compatibility with \texttt{Open Babel}~\cite{oboyle2011openbabel}. While core installation via \code{pip} requires the manual setup of external scientific dependencies, including \texttt{OpenMM}~\cite{eastman2023openmm8}, \texttt{PDBFixer}~\cite{eastman2017openmm7}, and \texttt{Vina}~\cite{Eberhardt2021Vina12}, Docker is fully supported as a seamless, reproducible alternative.

\subsection{Environment configuration and directory architecture}
\label{sec:si-layout}

\prodock uses a split-configuration approach, keeping project-level settings entirely separate from receptor and ligand definitions. This design makes the inputs easier to read and significantly simplifies reusing components across different campaigns. For instance, one can easily run the same set of receptors against a brand-new ligand library, or test the same ligands using different docking engines, all without rewriting the entire workflow.

Here is what a typical project directory looks like before running a campaign:

\begin{lstlisting}[language={},caption={Typical project layout before execution.}]
project/
|-- config.json
|-- receptor.json
|-- ligand.json
\end{lstlisting}

As the campaign runs, \prodock automatically generates a serialized campaign summary and organizes the preparation and result files into receptor-specific folders. A representative project layout after a successful run is shown below:

\begin{lstlisting}[language={},caption={Representative project layout after execution. Additional preparation artifacts may be present depending on the workflow mode.}]
project/
|-- campaign.json
|-- config.json
|-- receptor.json
|-- ligand.json
|-- <receptor_id>/
|   |-- filtered_protein/
|   |-- receptor/
|   |-- reference_ligand/
|   |-- results/
|       |-- docked/
|       |-- logs/
|-- prodock.db
\end{lstlisting}

Maintaining all associated files within a unified project directory is a foundational element of the reproducibility model in \prodock. By co-locating the original inputs, derived preparation artifacts, docking outputs, execution logs, and the optional SQLite database, the complete state of a computational campaign is natively preserved. This architecture allows previous experiments to be analyzed without external bookkeeping. Additionally, the auto-generated \code{campaign.json} records the exact runtime parameters, providing a permanent historical snapshot even if the primary receptor or ligand files are subsequently reused.

To illustrate this structure, an abstract overview of the three JSON inputs is shown below:

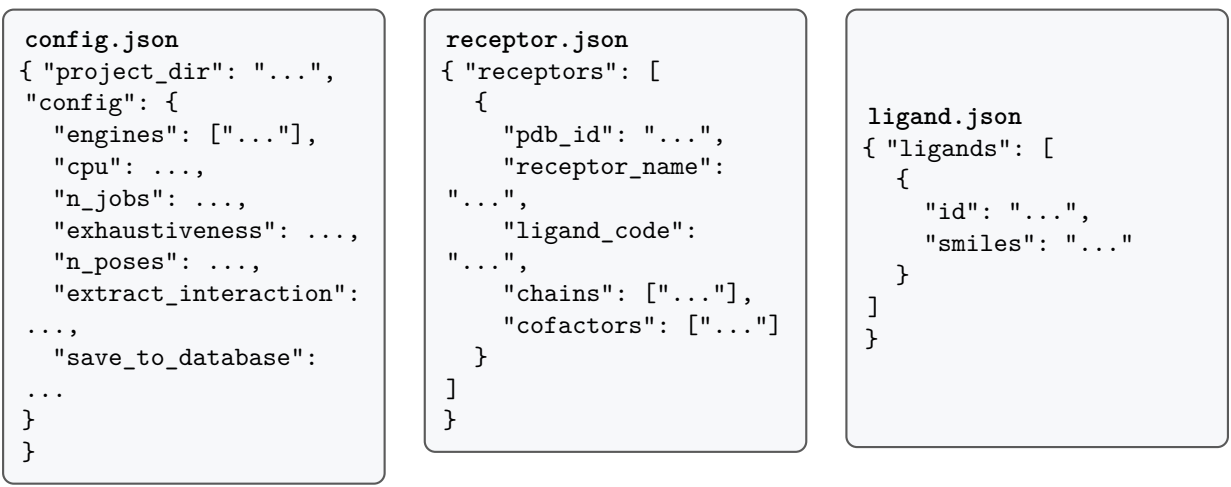
\begin{figure}[htbp]
\centering
\begin{tikzpicture}[
    font=\small\ttfamily,
    box/.style={
        draw=jsonline,
        rounded corners=4pt,
        thick,
        fill=jsonbg,
        align=left,
        inner sep=8pt,
        text width=0.28\linewidth,
        minimum height=5.8cm,
        anchor=north
    }
]

\node[box] (config) at (0,0) {
\textbf{config.json}

\{
  "project\_dir": "...",\\
  "config": \{\\
  \quad "engines": ["..."],\\
  \quad "cpu": ...,\\
  \quad "n\_jobs": ...,\\
  \quad "exhaustiveness": ...,\\
  \quad "n\_poses": ...,\\
  \quad "extract\_interaction": ...,\\
  \quad "save\_to\_database": ...\\
  \}\\
\}
};

\node[box, right=0.5cm of config.north east, anchor=north west] (receptor) {
\textbf{receptor.json}

\{
  "receptors": [\\
  \quad \{\\
  \quad\quad "pdb\_id": "...",\\
  \quad\quad "receptor\_name": "...",\\
  \quad\quad "ligand\_code": "...",\\
  \quad\quad "chains": ["..."],\\
  \quad\quad "cofactors": ["..."]\\
  \quad \}\\
  ]\\
\}
};

\node[box, right=0.5cm of receptor.north east, anchor=north west] (ligand) {
\textbf{ligand.json}

\{
  "ligands": [\\
  \quad \{\\
  \quad\quad "id": "...",\\
  \quad\quad "smiles": "..."\\
  \quad \}\\
  ]\\
\}
};

\end{tikzpicture}
\caption{Abstract structure of the three JSON files used in split-configuration mode. Project-level runtime settings are stored separately from receptor and ligand definitions.}
\label{fig:split-json-schema}
\end{figure}

Overall, this split-JSON design improves both clarity and reuse. It provides a simple interface for routine execution, while still supporting more iterative use cases in which receptors, ligands, or campaign settings are exchanged independently across repeated docking experiments.

\section{Autobox construction algorithms in \prodock}
\label{sec:si-gridbox}

The \class{GridBox} module within \prodock is responsible for constructing docking boxes based on reference-ligand coordinates. Recognizing that no single sizing strategy accommodates every molecular shape, the implementation provides multiple autobox algorithms. These range from practical defaults for standard applications to specialized methods for elongated ligands, outlier-robust boundary estimation, and multi-reference consensus boxes. Underlying coordinate operations are exclusively managed by \texttt{RDKit}~\cite{rdkit}.

\begin{table}[htbp]
\centering
\caption{Autobox algorithms currently exposed by \class{GridBox}.}
\label{tab:si-gridbox}
\small
\begin{tabularx}{\linewidth}{@{}l X X@{}}
\toprule
\textbf{Algorithm} & \textbf{Description} & \textbf{Typical parameters} \\
\midrule
\code{scale} & Multiplies the ligand span along each axis by a scale factor; useful as a fast default. & \code{scale}, \code{isotropic} \\
\code{pad} & Adds symmetric padding to the ligand span and can optionally enforce a minimum box size. & \code{pad}, \code{min\_size}, \code{isotropic} \\
\code{advanced} & Padding-based construction with optional heavy-atom filtering and optional snapping or rounding. & \code{pad}, \code{heavy\_only}, \code{snap}, \code{round\_ndigits} \\
\code{percentile} & Uses coordinate quantiles rather than extrema to reduce the effect of outlying atoms. & \code{low}, \code{high}, \code{pad}, \code{isotropic} \\
\code{pca-aabb} & Constructs an oriented envelope in principal-component space and returns the enclosing axis-aligned box. & \code{pca\_scale}, \code{pca\_pad}, \code{round\_ndigits} \\
\code{centroid-fixed} & Centers the box at the ligand centroid and applies a user-defined box size. & \code{size} \\
\bottomrule
\end{tabularx}
\end{table}

Table~\ref{tab:si-gridbox} outlines the available sizing algorithms: \code{pad} and \code{advanced} serve as practical defaults for most small molecules, whereas \code{pca-aabb} is tailored for elongated ligands. To prevent rare coordinate outliers from inflating the box dimensions, the \code{percentile} mode can be used. By running autobox generation as a formal preprocessing step, \prodock automatically captures the resulting box dimensions as reproducible workflow artifacts, replacing the manual step of defining them in a separate 3D viewer.

\section{Database schema and query examples}
\label{sec:si-db}

\begin{figure}[htbp]
\centering
\IfFileExists{Figure/database_schema.png}{%
    \includegraphics[width=0.9\linewidth]{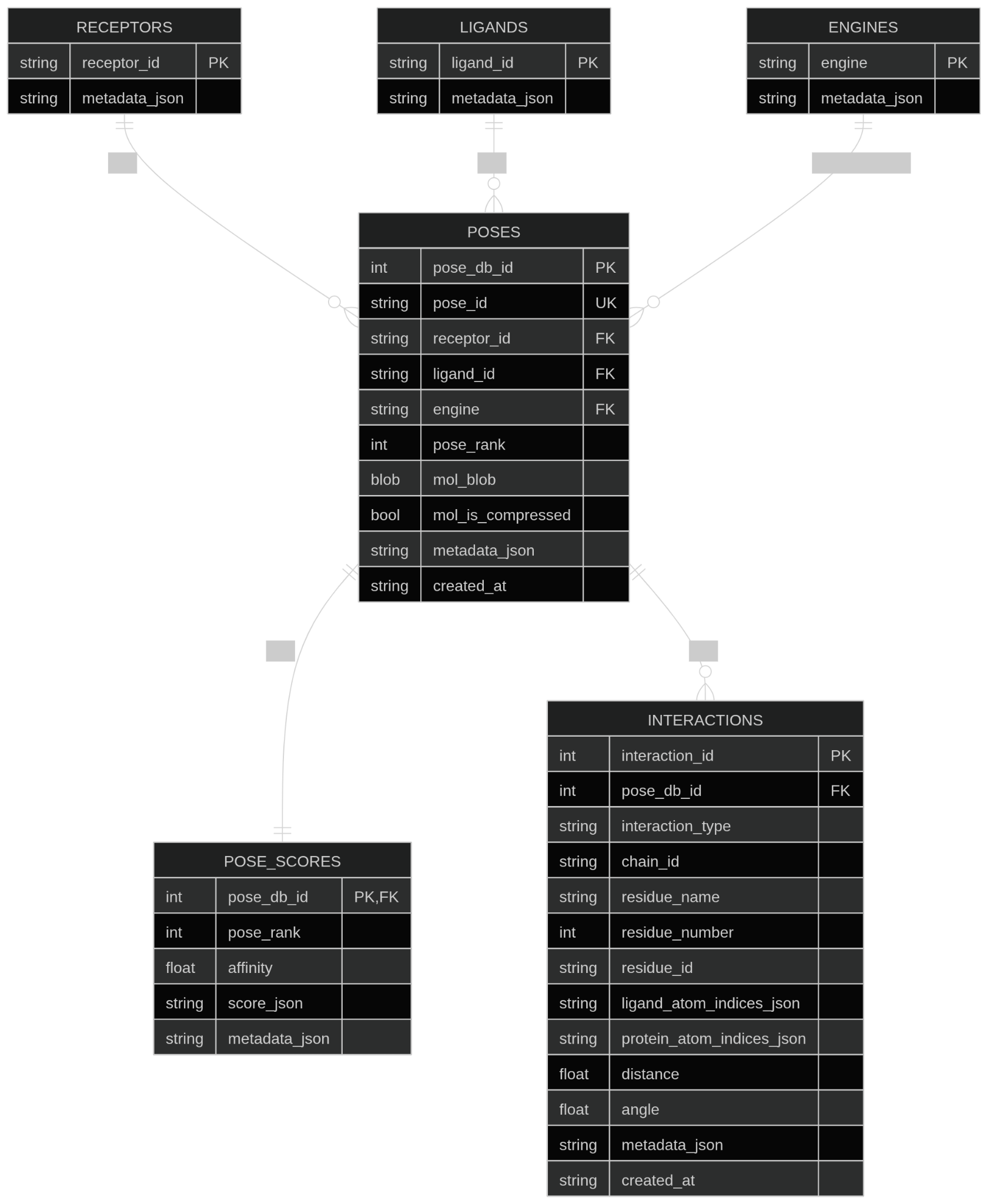}%
}{%
    \IfFileExists{../Figure/database_schema.png}{%
        \includegraphics[width=0.9\linewidth]{../Figure/database_schema.png}%
    }{%
        \fbox{\parbox[c][5.5cm][c]{0.9\linewidth}{\centering
        Placeholder for database schema figure.}}%
    }%
}
\caption{Database schema used by the \prodock SQLite layer. Dimension tables manage receptors, ligands, and engines, while factual tables capture docking geometries, scores, and residue-level interactions.}
\label{fig:si-db-schema}
\end{figure}

\prodock stores structured outputs in a normalized SQLite database. The schema employs dimension tables for receptors, ligands, and engines, alongside dedicated tables for poses, scores, and interactions. This design ensures that both the docking results and their contextual metadata are meticulously preserved, while keeping the database footprint small enough for seamless project archival. A simplified overview of the schema is provided in Figure~\ref{fig:si-db-schema}.

The core database entities are summarized in Table~\ref{tab:si-db-entities}.

\begin{table}[htbp]
\centering
\caption{Core entities in the \prodock SQLite schema.}
\label{tab:si-db-entities}
\small
\begin{tabularx}{\linewidth}{@{}lX@{}}
\toprule
\textbf{Entity} & \textbf{Role} \\
\midrule
\code{receptors} & Receptor identifiers and metadata. \\
\code{ligands} & Ligand identifiers and metadata. \\
\code{engines} & Docking-engine identifiers and metadata. \\
\code{poses} & Pose records linking receptor, ligand, and engine, with rank and serialized structures. \\
\code{pose\_scores} & Pose-level affinity values and optional score fields. \\
\code{interactions} & Residue-level interaction records linked to poses. \\
\bottomrule
\end{tabularx}
\end{table}

\clearpage

A representative programmatic query is demonstrated below:

\begin{lstlisting}[language=Python,caption={Example analytical query utilizing \class{PoseQuery}.}]
from prodock.database import PoseQuery

q = PoseQuery("project/prodock.db")

poses = q.poses(
    receptor_id="4ZAU",
    engine="qvina",
    as_dataframe=True,
)

hydrophobic = q.interactions(
    receptor_id="4ZAU",
    interaction_type="Hydrophobic",
    as_dataframe=True,
)

fp = q.fingerprint(
    receptor_id="4ZAU",
    mode="binary",
    index_by="pose_key",
)
\end{lstlisting}

This query architecture significantly strengthens the overall workflow. By consolidating data post-execution, complex analytical tasks and data extraction can be performed directly against the SQLite database, entirely replacing the inefficient practice of repetitively parsing raw engine outputs.

\section{End-to-end workflow examples}
\label{sec:si-usage}

\prodock supports both Python scripting and command-line execution through the same workflow, including receptor and ligand preparation, docking, postprocessing, and optional database storage. The Python API is well suited for interactive exploration, whereas the command-line interface (CLI) supports configuration-driven and reproducible project execution.

A minimal end-to-end workflow using the Python interface is shown below.

\begin{lstlisting}[language=Python,caption={Illustrative end-to-end workflow using the Python interface.}]
from prodock import prodock

result = prodock(
    project_dir="project",
    receptors=[
        {
            "pdb_id": "4WKQ",
            "ligand_code": "IRE",
            "chains": ["A"],
        }
    ],
    ligands=[
        {
            "id": "erlotinib",
            "smiles": "COCCOc1cc2c(ncnc2cc1OCCOC)Nc1cccc(c1)C#C",
        }
    ],
    engines=["qvina", "vina"],
    extract_interaction=True,
    save_to_database=True,
)

print(result.campaign_json)
print(result.db_path)
\end{lstlisting}

In this example, a single Python call prepares the receptor and ligand inputs, defines the docking campaign, executes the selected docking engines, postprocesses the resulting poses, and stores the generated outputs in a self-contained project directory. The returned object provides direct references to key artifacts, including the serialized campaign file and the SQLite database.

The same workflow can also be executed from the command line.

\begin{lstlisting}[language=bash,caption={Illustrative end-to-end workflow using the command-line interface.}]
python -m prodock \
  --config project/config.json \
  --receptor-json project/receptor.json \
  --ligand-json project/ligand.json
\end{lstlisting}

In the CLI workflow, global execution settings such as docking engines, runtime parameters, and database options are stored in \code{config.json}, whereas receptor and ligand definitions are provided separately through \code{receptor.json} and \code{ligand.json}. This separation simplifies reuse of the same molecular inputs across multiple docking campaigns with different execution settings.

Although the Python and CLI entry points differ in usage style, both invoke the same underlying workflow and generate the same structured project outputs. This unified design allows \prodock to support both rapid interactive analysis and larger reproducible screening studies.

\section{Case study}
\label{sec:si-case-study}

To illustrate \prodock in practice, we simulated an EGFR docking campaign against five receptor structures (\texttt{2ITY}, \texttt{1M17}, \texttt{4G5J}, \texttt{4I23}, and \texttt{4ZAU}). Our ligand panel consisted of five established EGFR inhibitors (\texttt{gefitinib}, \texttt{erlotinib}, \texttt{afatinib}, \texttt{dacomitinib}, and \texttt{osimertinib}) alongside 20 additional compounds to diversify the test set. Rather than serving as a formal virtual screening benchmark, this case study demonstrates how \prodock seamlessly orchestrates a docking study involving multiple receptors, ligands, and engines into a single reproducible workflow.

Starting with only PDB identifiers and ligand \texttt{SMILES} strings, \prodock fully automated the system preparation. For the receptors, it retrieved the source structures, selected specific chains, stripped solvent molecules, retained necessary cofactors, and ultimately generated ready-to-dock \texttt{PDBQT} files. Simultaneously, \prodock embedded the ligands in three dimensions, optimized their geometries, and converted them into \texttt{PDBQT} format within a centralized directory. This shared setup allowed the prepared ligand library to be efficiently reused across all combinations of receptors and engines.

To reflect a typical structural docking scenario, we opted to define search boxes automatically using cocrystallized reference ligands rather than manual coordinates. Specifically, \prodock derived the docking grids from the native ligands \texttt{IRE}, \texttt{AQ4}, \texttt{0WM}, \texttt{1C9}, and \texttt{YY3} (for receptors \texttt{2ITY}, \texttt{1M17}, \texttt{4G5J}, \texttt{4I23}, and \texttt{4ZAU}, respectively). This approach keeps box generation transparent and tightly integrated with the rest of the structural preprocessing.

Following preparation, \prodock compiled the setup into a serialized campaign file and executed the docking runs across four backends: \texttt{smina}, \texttt{vina}, \texttt{qvina}, and \texttt{qvinaw}. We applied uniform parameters across all jobs: four CPUs per job, four parallel executions, an exhaustiveness of 32, and 20 requested poses per ligand. Finally, \prodock organized the outputs into a neat receptor-focused project directory containing the prepared files, logs, docked poses, and a local SQLite database for straightforward downstream analysis.

Ultimately, this case study serves as a practical demonstration of the \prodock lifecycle. It shows that the entire pipeline can be handled within a cohesive framework. From retrieving raw molecular inputs and mapping binding sites to parallel batch execution and structured data collection, the full process is unified. Rather than acting as a simple wrapper for a single tool, \prodock delivers a fully reproducible and comprehensive workflow that transforms raw inputs into clean data ready for analysis.

\bibliographystyle{unsrt}
\bibliography{reference}

\end{document}